\documentclass[5p]{elsarticle}

\usepackage[colorlinks,hyperindex]{hyperref}
\usepackage[utf8x]{inputenc}
\usepackage{float}
\usepackage{latexsym}
\usepackage{graphicx}
\usepackage{amssymb}
\usepackage{amsmath}
\usepackage{braket}

\begin{document}

\title{Uncover quantumness in the crossover from  BEC  to  quantum-correlated phase}

\author{J. P. Restrepo Cuartas\corref{cor1}}
\ead{jprestrepocu@unal.edu.co}
\author{H. Vinck-Posada}

\address{Departamento de F\'isica, Universidad Nacional de Colombia, 111321, Bogot\'a, Colombia}

\cortext[cor1]{Corresponding author}

\date{\today}

\begin{abstract}
Collective phenomena in the  Tavis-Cummings model has been widely studied, focusing on the phase transition features. In many occasions, it has been used variational approaches that consider separated radiation-matters systems. In this paper, we examine the role of the quantum entanglement of an assembly of two-level emitters coupled to a single-mode cavity; this allows us to characterise the quantum  correlated state for each regime. Statistical properties of the system, e.g., the first four statistical moments, show clearly the structure of the light and matter distributions. Even though that the second order correlation function goes to one in some regimes, the statistical analysis evidence a sharp departure from coherent behaviour, contrarily to the common understanding. 
\end{abstract}

\maketitle

\section{Introduction}\label{intro}
\noindent
Quantum correlations (QC) constitute a key component
of the successful development of quantum technologies~\cite{Mooney19,delvalle07b,RestrepoCuartas15}.
Nowadays, several groups and industries are involved in the
run for the experimental realisation, e.g.,
the race for a successful ‘quantum supremacy’ 
breakthrough~\cite{Arute19}, quantum sensing~\cite{Zhou20},
and the  use of squeezed vacuum states in the direct
measurement of gravitational waves~\cite{Tse19}. In this sense, a lot of promising systems had been studied as candidates, e.g., quantum dots in semiconductor microcavities~\cite{Najer19}, all-optical qubits~\cite{OBrien03}, superconductor circuits~\cite{Stassi20,Arute19}, ion-traps~\cite{Georgescu20}, and atomic clusters~\cite{Choi19}. 

Entanglement was widely addressed in recent years. Eigensystem approach is one of the main methods used to describe the stationary state of the Dicke assembly coupling. This approach focuses on the correlation analysis of the ground state without taking into account the light field relevance~\cite{Rodriguez-Lara10,Robles15,Mao16c}. Recently, an outstanding experimental realisations is featured. Dicke model comes to the lab reality~\cite{Quiroz-Juarez20}. 

Strong correlations in an assembly of qubits 
has been addressed in the seminal works about Bose-Einstein 
condensation of cavity polaritons~\cite{senellart99a, lesidang98a}. A broad interest in the crossover from the
quasi-condensed to the quantum-correlated plasma at high
densities where a  description of  the ground state to
get the main features of the Bose-Einstein condensate to Quantum correlated phase transition (BEC-QCP)~\cite{eastham01a,Yamaguchi13,Horikiri16, byrnes10a,Chen05}. Therein, the authors have used 
variational approaches taking into account a pure trial
state built up by the uncorrelated product of a BCS-like
state and a light coherent state. Besides this kind
of approaches, which most likely addresses classical
correlations ---lack of entanglement and other quantum resources--- another point of view that focuses on
actual quantum correlations have received a lot of
attention in recent years~\cite{Mao16}. Avoiding the 
separability of the total quantum state, it is  necessary
to account for the quantum correlations which arise
from the light-matter  interaction.

In this article, we use the Tavis-Cummings
model~\cite{dicke54a,tavis68a} to obtain the stationary 
behaviour close to the classical limit. First, using an effective
atomic coherent state, i.e, $\ket{\theta,\varphi}$~\cite{Yamamoto99}, we make
a comparison with previous works where,  once
again, it
is considered a separable global state of
light and matter $\ket{\theta,\varphi}\otimes\ket{\alpha}$.
Then,  to go beyond this, we use an exact diagonalisation
approach consequently, with the eigenstates of each
excitation manifold, we are able to catch up the entanglement
features of the  crossover. Finally, we identify the
statistical properties for the polaritons and characterise their relationship with the phase transition. 

The paper is organised as follows: Section 2 introduces the theoretical model. Section 3  present the results, and in section 4 summarise and conclude. 

\section{Theoretical model}\label{sec:formalism}

\begin{figure*}[ht]
\centering
\includegraphics[width=0.9\textwidth]{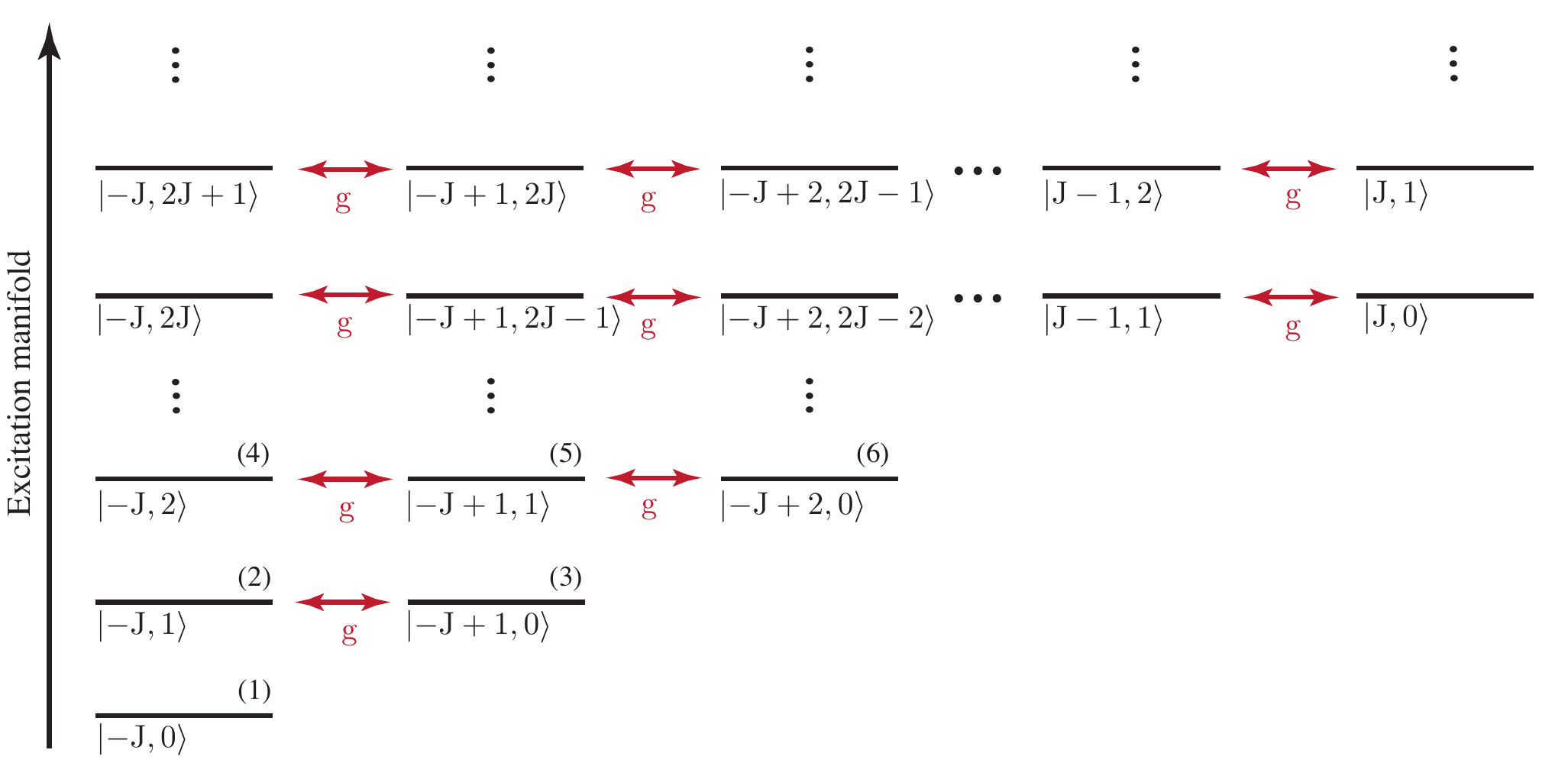}
\caption{Ordering of states  $\ket{\mathrm{M,n}}$ by excitation manifold. The first index is the angular momentum quantum number, and the second one is the number of photons. The ordering is described in the main text in equation~(\ref{eq:manif}).
}\label{fig:dickebasis}
\end{figure*}

The theoretical model involves an assembly
of quantum emitters (QEs) coupled to a single cavity mode
through the TC Hamiltonian that was successful for describing multiple experimental realisations as it is shown schematically in Fig \ref{fig:dickebasis} .
The well known TC Hamiltonian which describes the system is
composed by a cavity mode with frequency $\omega_c$,
the N two-level atoms with frequencies
$\omega_{a}=\omega_c-\Delta$, and the 
light--matter coupling due to a linear dipolar
interaction can be written as
\begin{equation}\label{eq:HTC}
      H_\mathrm{TC}= \omega_{\mathrm{c}} a^\dag a + \omega_\mathrm{a}{J_\mathrm{z}}+\frac{\mathrm{g}}{\sqrt{\mathrm{N}}}(a^\dag J_ - + a J_+
      ),
\end{equation}
where $ a^\dag$ ($ a$) is the bosonic creation
(annihilation) operator in the Fock basis, where $\hbar=1$. $ J_z,
J_{\pm}$  are collective angular momentum operators for the set of N two-level emitters, and
$\mathrm{g}$ is the matter--field strength 
coupling \cite{garraway11}.
\subsection{Angular momentum states}%
To describe the state of the matter component for the system, it is necessary to define the set of two-level operators ($\sigma^{\mathrm{i}}_\mathrm{z}$, $\sigma^{\mathrm{i}}_{\pm}$), which can be written in terms of the matter bare basis ($\ket{\mathrm{g}_{i}}, \ket{\mathrm{e}_{i}}$), as
\begin{align}\label{eq:SigmaOps}
       \sigma_\mathrm{z}^\mathrm{i}&=\ket{\mathrm{e}_\mathrm{i}}\bra{\mathrm{e}_\mathrm{i}}-\ket{\mathrm{g}_\mathrm{i}}\bra{\mathrm{g}_\mathrm{i}},\,\,\sigma_+^\mathrm{i}=\ket{\mathrm{e}_\mathrm{i}}\bra{\mathrm{g}_\mathrm{i}},\,\, \text{and} \,\,\\ \nonumber \sigma_-^\mathrm{i}&=\ket{\mathrm{g}_\mathrm{i}}\bra{\mathrm{e}_\mathrm{i}}
\end{align}
where the exterior product $\ket{\Psi}=\bigotimes^{\mathrm{N}}_{\mathrm{i}}\ket{\alpha_{\mathrm{i}}}$ form the $\mathrm{N}$-emitters Hilbert space, with $\alpha_{\mathrm{i}}=\mathrm{g}_{\mathrm{i}},\mathrm{e}_{\mathrm{i}}$, and the operators must be understood as $\sigma^{\mathrm{i}}_{\mathrm{j}}:=\dotsm \otimes\mathbb{I}^{\mathrm{i}-1}\otimes\sigma^{\mathrm{i}}_{\mathrm{j}}\otimes\mathbb{I}^{\mathrm{i}+1}\otimes\dotsm$, and $\mathrm{j}=\mathrm{z},\, \pm$.
The collective angular momentum operators of the TC Hamiltonian in Eq. (\ref{eq:HTC}) can be written in terms of the $\mathrm{N}$-emitters operators as follows
\begin{equation}
       J_{\mathrm{z}}=\frac{1}{2}\sum_{\mathrm{i}=1}^{\mathrm{N}}{\sigma_\mathrm{z}^\mathrm{i}}\,\,\,\, \text{and} \,\,\,\,  J_{\mathrm{\pm}}=\frac{1}{2}\sum_{\mathrm{i}=1}^{\mathrm{N}}{\sigma_\mathrm{\pm}^\mathrm{i}}.
      \label{eq:JOps}
\end{equation}
The set of angular momentum operators 
satisfy the commutation relations
$[ J_+, J_-]=2 J_\mathrm{z}$. The eigenstates of $J^{2},\,J_{\mathrm{z}}$ as it is well-known in the literature, the
Dicke basis \cite{dicke54a,Arecchi72,Yamamoto99}, can be generated from the vacuum state  ($J_-\ket{-\mathrm{J}}=0$),
that takes the form
\begin{equation}
      \ket{\mathrm{M}}:=\ket{\mathrm{J},\mathrm{M}}=\frac{1}{(\mathrm{M}+\mathrm{J})!}\binom{2\mathrm{J}}{ \mathrm{M}+\mathrm{J}}^{-\frac{1}{2}}J_+^{\mathrm{M}+\mathrm{J}}\ket{-\mathrm{J}},
      \label{DickeState1}
\end{equation}
bear in mind that the eigenvalues of $J^{2}$ and $J_{\mathrm{z}}$ are $\mathrm{J}(\mathrm{J}+1)$ and $\mathrm{M}$, respectively, with
$\mathrm{M}=-\mathrm{J},-\mathrm{J}+1,\dots,\mathrm{J}$. Moreover, the non-Hermitian operator $J_{+}$ ($J_{-}$) increases (decreases) the eigenvalue $\mathrm{M}$ to $\mathrm{M}+1$ ($\mathrm{M}-1$). For a fixed number $\mathrm{N}$ of two-level systems, we will suppress the quantum number $\mathrm{J=N/2}$ from the state notation. The raising/lowering operators satisfy the usual relation
\begin{equation}J_{\pm}\ket{\mathrm{M}}=\sqrt{\mathrm{J(\mathrm{J}+1)}-\mathrm{M}(\mathrm{M}\pm 1)}\ket{\mathrm{M}\pm 1}.
\end{equation}

%\end%%

\subsection{Atomic coherent
states}%

\begin{figure}
\begin{centering}
\includegraphics[scale=0.3]{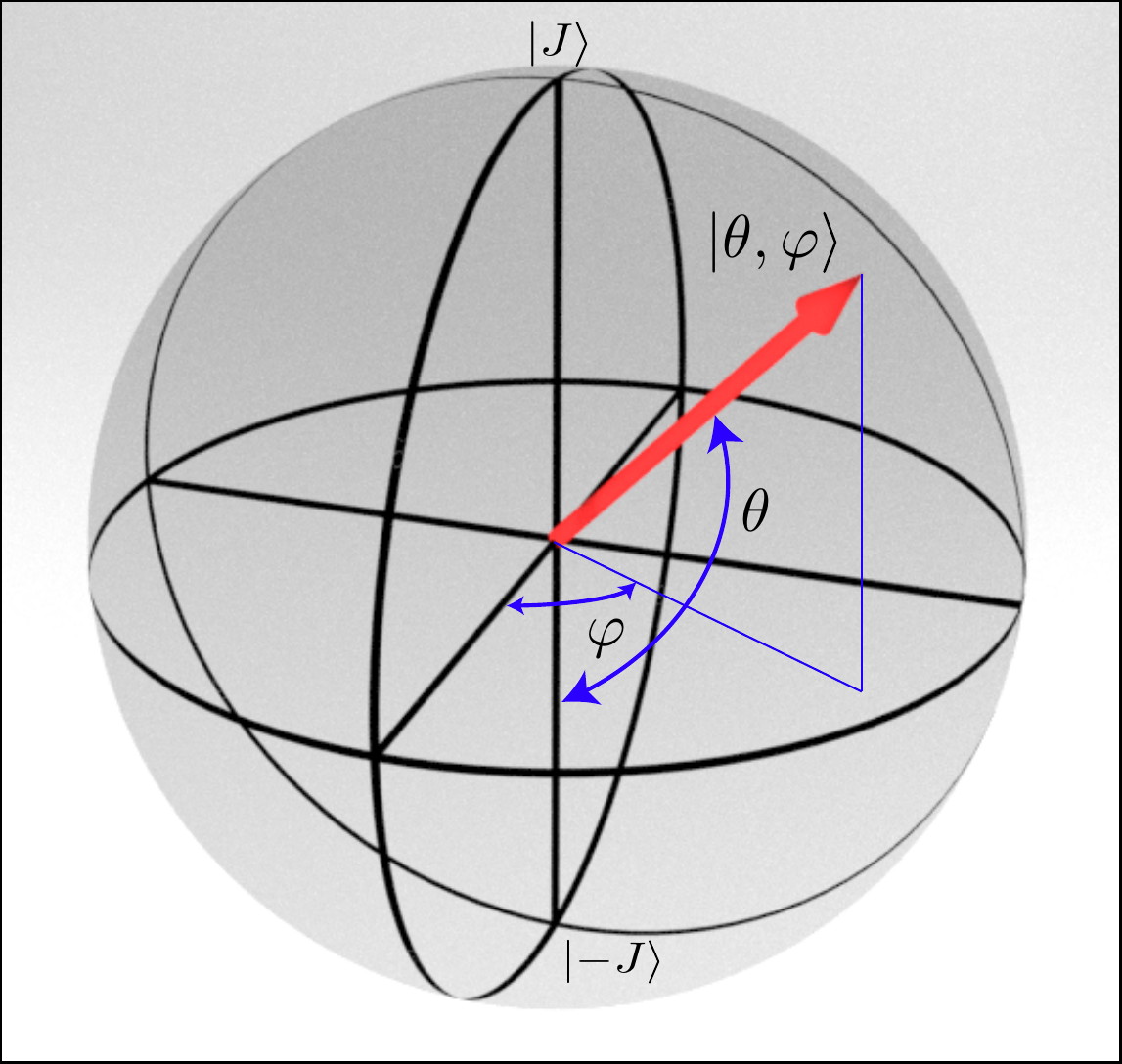}
\caption{Rotation on the Bloch sphere surface by the application of the rotation operator $\mathcal{R_{\theta,\varphi}}$ over the ground state $\ket{-J}$ \cite{Arecchi72}.}
\label{BS}
\end{centering}
\end{figure}

To build up an atomic coherent state, also
known as a Bloch state, $\ket{\theta,\varphi}$,
it is required to define a rotation operator, 
namely $\mathcal{ R_{\theta,\varphi}}$, which 
generate a rotation in $\theta$ and $\varphi$ 
on the surface of the Bloch sphere when acting 
over the ground state $\ket{-\mathrm{J}}$ 
(see Fig. \ref{BS}), such that
\begin{equation}
      \ket{\theta,\varphi}=\mathcal{ R_{\theta,\varphi}}\ket{-\mathrm{J}},
      \label{DickeState2}
\end{equation}
where the rotation operator $\mathcal{ R_{\theta,\varphi}}$ has the form
\begin{equation}
      \mathcal{ R_{\theta,\varphi}}=e^{\zeta J_+ - \zeta^* J_-},
      \label{DickeState3}
\end{equation}
where $\zeta=\frac{1}{2}\theta e^{i\varphi}$.
In this way, the action of the rotation
operator $\mathcal{ R_{\theta,\varphi}}$ over
the Dicke ground state leads to
\begin{equation}
      \ket{\theta,\varphi}=\mathcal{ R_{\theta,\varphi}}\ket{-\mathrm{J}}=\frac{1}{1+|\tau|^2}e^{\tau  J_+}\ket{-\mathrm{J}},
      \label{DickeState4}
\end{equation}

here $\tau=e^{-i \varphi}\tan{\theta /2}$.

The completeness relation is

\begin{equation}
    \int d\Omega \ket{\theta,\varphi}\bra{\theta,\varphi}=\frac{2\mathrm{J}+1}{4\pi} \label{eq:completeness}
\end{equation}

Using this expression, we can write the span 
of any arbitrary state in the coherent state basis

\begin{equation}
    \ket{\Psi}=\int d\Omega\, C(\theta,\varphi)\ket{\theta,\varphi} \label{eq:arbket}
\end{equation}
where 
\begin{align}
      C(\theta,\varphi)&=\left(\frac{4\pi}{2\mathrm{J}+1}\right)^{\frac{1}{2}}\braket{\theta,\varphi|\Psi}\\ \nonumber
      &=\left[\frac{4\pi}{2\mathrm{J}+1}\binom{2\mathrm{J}}{\mathrm{M}+\mathrm{J}}\right]^{\frac{1}{2}}\sin^{\mathrm{J}+\mathrm{M}}\left(\frac{\theta}{2}\right)\cos^{\mathrm{J}-
      \mathrm{M}}\left(\frac{\theta}{2}\right)\\ \nonumber
      &\times\exp(i(\mathrm{J}+\mathrm{M})\varphi)
      \label{eq:coeff}
\end{align}

Now, to study a phase transition in the system, the interest is to find out the optimal $\theta$ and $\varphi$ values such that minimise the expectation value of the Hamiltonian restricted to a fixed number of particles 
%%%%%%
\begin{equation}
    \left<\Bar{M}\right> =\left<{H}-\mu{N}\right>
\end{equation}
%%%%%
\begin{align}
    \nonumber\Bar{M}=&(\omega_{c}-\mu)\alpha\alpha^{*}-(\omega_{\textrm{a}}-\mu)\mathrm{J}\cos{\theta}-\mu \mathrm{J}\\
    \nonumber &+\mathrm{g}\sqrt{\frac{\mathrm{J}}{2}}\sin{\theta}(\alpha e^{i\varphi}+\alpha^{*}e^{-i\varphi})\\
    \nonumber &+\mathrm{g}\eta\sqrt{\frac{\mathrm{J}}{2}}\sin{\theta}(\alpha e^{-i\varphi}+\alpha^{*}e^{i\varphi})\\
    &+\epsilon\sqrt{8\mathrm{J}^{3}}\sin{\theta}\cos{\varphi}.
\end{align}
%%%%%
Taking $\alpha=\alpha^{*}$,
%%%%%
\begin{align}
    \nonumber \Bar{M}=&(\omega_{c}-\mu)\alpha^{2}-(\omega_{\textrm{a}}-\mu)\mathrm{J}\cos{\theta}-\mu \mathrm{J}\\
    &\!+\left[\mathrm{g}\sqrt{2\mathrm{J}}(1+n)\alpha+\epsilon\sqrt{8\mathrm{J}^{3}}\right]\sin{\theta}\cos{\varphi}
\end{align}
%%%%s
The minimisation of $\Bar{M}$ respect to the parameters $\alpha$, $\theta$, $\varphi$ yields 
\begin{subequations}
\begin{align}
    \frac{\partial{\Bar{M}}}{\partial\alpha}&=2(\omega_{c}-\mu)\alpha+\mathrm{g}\sqrt{2\mathrm{J}}(1+n)\sin{\theta}cos{\varphi}=0.\\ 
    \nonumber\frac{\partial{\Bar{M}}}{\partial\theta}&=(\omega_{\textrm{a}}-\mu)\mathrm{J}\sin{\theta}\\
    &\!+\left[\mathrm{g}\sqrt{2\mathrm{J}}(1+n)\alpha+\epsilon\sqrt{8\mathrm{J}^{3}}\right]\cos{\theta}\cos{\varphi}=0.\\
    \frac{\partial{\Bar{M}}}{\partial\varphi}&=-\left[\mathrm{g}\sqrt{2\mathrm{J}}(1+n)\alpha+\epsilon\sqrt{8\mathrm{J}^{3}}\right]\sin{\theta}\sin{\varphi}=0
    \label{MinPhi}
\end{align}
\end{subequations}
%%%%
Now, solving for the values of the parameters that minimise the quantity $\Bar{M}$
%%%%
\begin{equation}
    \sin{\theta=-\frac{2(\omega_{c}-\mu)\alpha}{\mathrm{g}\sqrt{2\mathrm{J}}(1+n)\cos{\varphi}}}
\end{equation}
%%%%
\begin{equation}
    \tan{\theta=\mp\frac{2(\omega_{c}-\mu)\alpha}{\left(2\mathrm{g}^{2}\mathrm{J}(1+n)^{2}\cos^{2}{\varphi}+4\left(\omega_{c}-\mu\right)^{2}\alpha^{2}\right)^{1/2}}}
\end{equation}
%%%%
\begin{align}
    \mp&\frac{\sqrt{2}\left(\mu-\omega_{c}\right)\left(\omega_{\textrm{a}}-\mu\right)\mathrm{J}\alpha}{\left(\mathrm{g}^{2}\mathrm{J}\left(1+n\right)\cos^{2}{\varphi}-2\left(\mu-\omega_{c}\right)^{2}\alpha^{2}\right)^{1/2}}\\ \nonumber
    =&\left[\mathrm{g}\sqrt{2\mathrm{J}}(1+n)\alpha+\epsilon\sqrt{8\mathrm{J}^{3}}\right]\cos{\varphi}.
\end{align}
%%%%
From Eq. (\ref{MinPhi}) $\varphi=0 \rightarrow \cos{(0)}=1$, this implies that
%%%%
\begin{align}
   \nonumber\pm\frac{\sqrt{2}\left(\mu-\omega_{c}\right)\left(\omega_{\textrm{a}}-\mu\right)J\alpha}{\left(\mathrm{g}^{2}\mathrm{J}\left(1+n\right)-2\left(\mu-\omega_{c}\right)^{2}\alpha^{2}\right)^{1/2}}&\\=\mathrm{g}\sqrt{2\mathrm{J}}(1+n)\alpha+\epsilon\sqrt{8\mathrm{J}^{3}}.
\end{align}

This system is solved along with the constraint that preserves the normalised excitation manifold, i.e., the mean value of the particle density operator  $N_{\mathrm{ex}}=\left( a^{\dagger} a +   J_{z}+\mathrm{J}\mathbb{I}\right)/\mathrm{N}$
is 
\begin{equation}
    \rho_{\mathrm{ex}}=\left< N_{\mathrm{ex}}\right>=\frac{1}{\mathrm{N}}\left(|\alpha|^{2}-\mathrm{J} \cos{\theta}+\mathrm{J}\right),
\end{equation}
%%%%
with $\mathrm{J}=\mathrm{N}/2$, so $\rho_{\mathrm{ex}}$ takes the form
%%%%
\begin{equation}
    \rho_{\mathrm{ex}}=\frac{|\alpha|^{2}}{2\mathrm{J}}-\frac{1}{2}\cos{\theta}+\frac{1}{2}\mathrm{J}.\end{equation}
%%%%

\subsection{Manifold Ground Dressed State Approach}

The Tavis-Cummings Hamiltonian commutes with the total excitation number operator $N=J_{\mathrm{z}} +a^\dagger a+\mathrm{J}$; therefore the N operator is a constant of motion. This fact entails that each excitation manifold is also a conserved quantity.

Consequently, the Hamiltonian becomes block-diagonal in the Dicke basis $\ket{\mathrm{M,n}}$ as it is organised in the figure~\ref{fig:dickebasis}. The block for the $\mathrm{n}$ manifold can be spanned in the following subsets:

\begin{subequations}\label{eq:manif}
\begin{align}
\mathrm{If}\,\,\, \mathrm{n<2J},\,\,\, &\mathrm{then}\,\,\,
\{\ket{\mathrm{-J,n}}, \ket{\mathrm{-J+1,n-1}},\\ \nonumber &\dots,\ket{\mathrm{-J+n-1,0}}, \ket{\mathrm{-J+n},0}\}.\\ 
 \mathrm{If}\,\,\, \mathrm{n=2J},\,\,\, &\mathrm{then}\,\,\, \{\ket{\mathrm{-J,n}}, \dots, \ket{\mathrm{2J-1},0}, \ket{\mathrm{2J},0}\}.\\
 \mathrm{If}\,\,\, \mathrm{n>2J},\,\,\, &\mathrm{then}\,\,\, \{\ket{\mathrm{-J,n}}, \ket{\mathrm{-J+1,n-1}},\\ \nonumber &\dots,\ket{\mathrm{2J-1,n-2J+1}}, \ket{\mathrm{2J,n-2J}}\}.
\end{align}
\end{subequations}

We organise the states in each manifold block by decreasing the number of photons. Therefore, we  can label the correspondent set with the initial photon number $\mathrm{n}$, i.e., the state with none matter excitation $\ket{\mathrm{-J,n}}$. The density of excitations $\rho_{\mathrm{\mathrm{ex}}}$ is related with the manifold number as $\rho_{\mathrm{ex}}=(\mathrm{n-J})/\mathrm{N}$. In this picture, the ground state corresponds to the lowest energy eigenstate for each manifold. Then, the chemical potential $\mu$ is calculated as the energy difference of the minimum energies of two successive manifolds $\mathrm{\mu(\rho_{ex})=\mu_n=E^0_{n+1}-E^0_{n}}$.  We diagonalise each excitation manifold and with the lower eigenstates we build up the total density matrix $\mathrm{\rho^0_T}=\ket{E^0_{n}}\bra{E^0_{n}}$ ---we are always restricted to the ground state, for this reason, we will avoid any related label on its state operator---. Then we obtain the reduced density matrix for the components $\mathrm{\rho_C=Tr_A[\rho_T]}$ and $\mathrm{\rho_A=Tr_C[\rho_T]}$. All the expectation values are calculated as usual $\braket{O}=Tr[O\rho_\mathrm{T}]$.  On the other hand, we desire a deep  characterising of the light state and its  actual departure of the photon statistics from the Poisson distribution (coherent behaviour). Therefore, we obtain the first four statistical moments, e.g., the mean number of photons $\lambda_1=\braket{a^\dagger a}=Tr[a^\dagger a\rho_{\mathrm{T}}]$, the variance or dispersion  $\lambda_2=\braket{(\Delta n)^2}=\braket{n^2}-\braket{n}^2=Tr[(\Delta n)^2\rho_{\mathrm{T}}]$ (where the dispersion operator reads as $\Delta n=n-\braket{n}$). Henceforth, high order statistical moments are defined as $\lambda_n=Tr[(\Delta n/\sqrt{\lambda_2})^n\rho_{\mathrm{T}}]$, particularly, the skewness $\lambda_3$ and the kurtosis $\lambda_4$.  As it is well known, to determine if a distribution is sub-Poissonian or super-Poissonian, we must compare the statistical moments, e.g., sub-(super-)Poissonian corresponds to  $\lambda_2<\lambda_1\,(\lambda_2>\lambda_1)$, respectively.

\section{Results}\label{Results}
\begin{figure}[ht]
\centering
\includegraphics[width=0.45\textwidth]{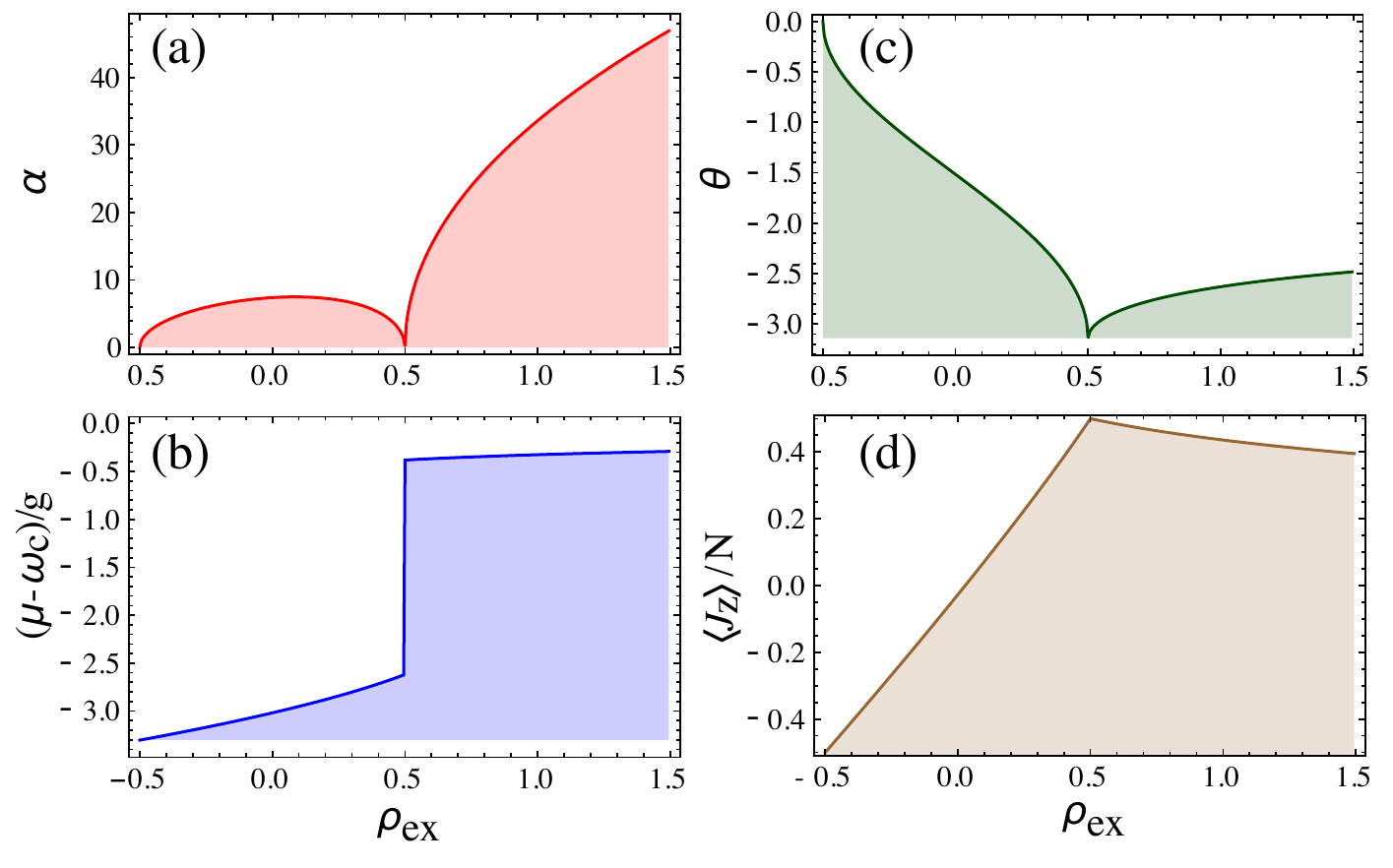}
\caption{Physical observables: (a) mean number of photons amplitude $\alpha$ (b) chemical potential $\mu$, (c) Excitation angle $\theta$, and (d) population inversion. All this quantities are calculated with the variational approach using the trial function  $\ket{\Psi}=\ket{\theta,\varphi}\otimes\ket{\alpha}$.}
\label{fig:fig3}
\end{figure}

\begin{figure*}[ht]
\centering
\includegraphics[width=0.9\textwidth]{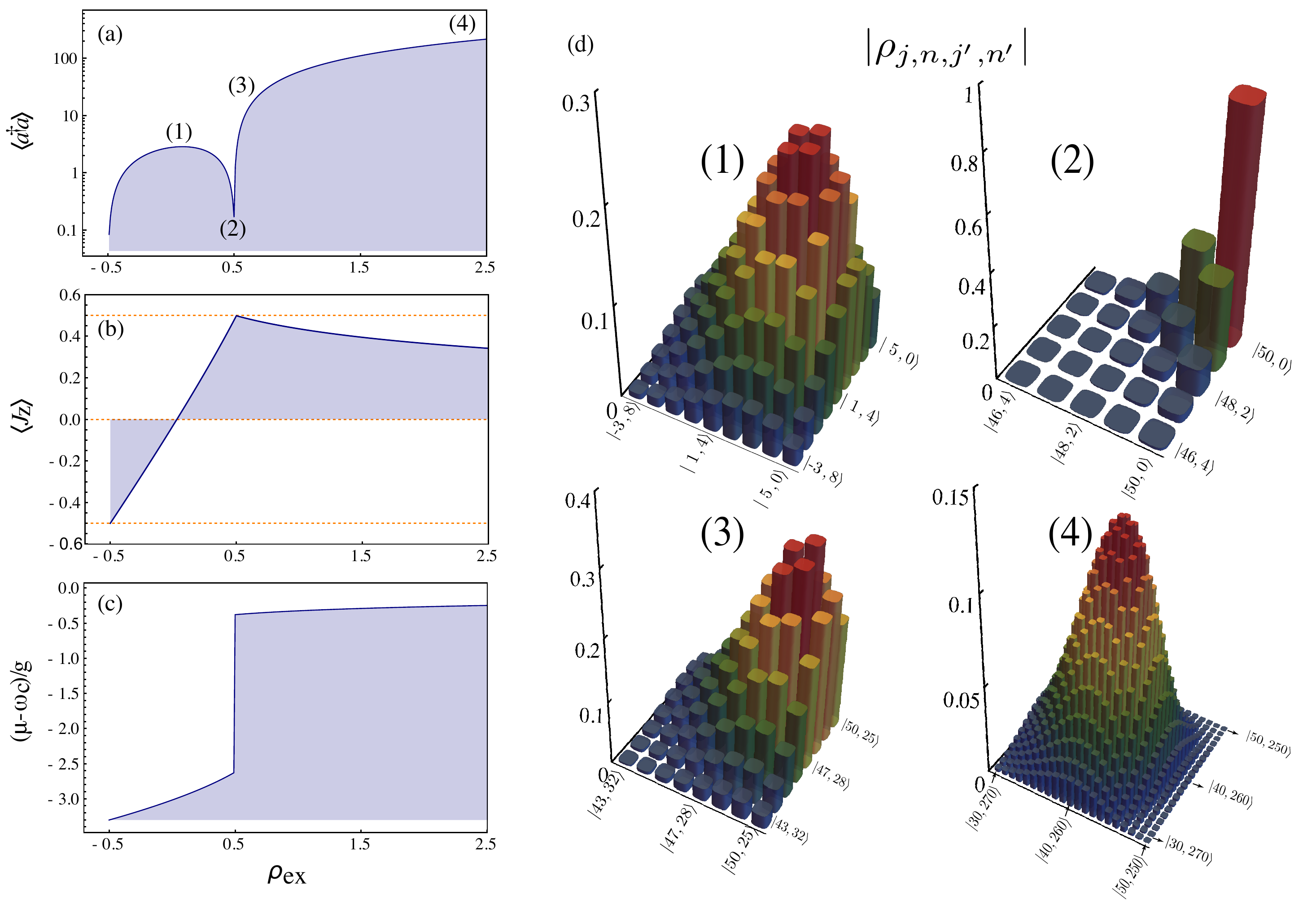}
\caption{Physical observables with the exact diagonalisation method. In panel (a) is the mean number of photons, (b) the population inversion, (c) the chemical potential, and (d)  the density matrix elements for four cases denoted into the panel (a). These correspond to values  before, on, and  after the crossover. The Hilbert space is denoted as $\ket{\textrm{M,n}}$}
\label{fig:tom}
\end{figure*}

At first, using a variational approach to model the system, similarly to the formalism used by~\cite{eastham01a}, but now adopting another kind of trial function: $\ket{\Psi}=\ket{\theta,\varphi}\otimes\ket{\alpha}$. Here, $\ket{\theta,\varphi}$ corresponds to a coherent state of an assembly of two level systems~\cite{Yamamoto99} and $\ket{\alpha}$ is a coherent state of light.  In figure \ref{fig:fig3} we show a) the behaviour of the amplitude  for the mean number of photons $(\alpha)$, b) the chemical potential $(\mu-\omega_c)/g$, c) the excitation angle of the matter coherent state $(\theta)$, and d) the normalised emitters population inversion $(\braket{J_\mathrm{z}}/\mathrm{N})$. All these results are in good agreement with the results reported by Easthman and Littlewood~\cite{eastham01a}, showing a well-known phase transition for $\rho_{ex}=0.5$ where the cavity is empty and whole matter assembly will be saturated. The main thing here, with this approach, is that it prevents a proper formation of entanglement between light and matter. 

Another way to focus this problem, bear in mind a description of the quantum entanglement features, can be built employing  an exact diagonalisation, identifying each block of conserved manifolds in the Hamiltonian. In this sense, a manifold is defined in concordance with equation~(\ref{eq:manif}). With this procedure, we ensure that quantum correlations between radiation and matter, in the sense of quantum entanglement, are well established.

Once again, after the exact diagonalisation process, the calculated observables give us similar information compared with the variational method. For $\Delta=3$, in figure \ref{fig:tom}(a) The mean number of photons is plotted (numbers indicate the order of density matrix elements depicted in part (d)). (b) shows the population inversion of two-level oscillators in the system. (c) shows the chemical potential, (d) the state represented by the density operator matrix. For the cases labelled as (1), (3), and (4), i.e., $\rho_{\mathrm{ex}}=\{0,0.75,2.5\}$ the states have an actual polaritonic structure with coherences densely populated. The mean number of photons for each case are around $2,26,\,\textrm{and}\,120$, respectively.  Other statistical moments shall be analysed shortly. For the case labelled as (2), i.e., $\rho_{\mathrm{ex}}=0.5$ the number of coherences is sharply reduced along as photons decrease, and the matter state saturates. This is evidence of radiation-matter decoupling.

Now, we analyse the quantum correlations. First, the second order correlation function, for light, suggest different regimes near the crossover. Before $\rho_{\mathrm{ex}}=0.5$, $g^2(0)$ increases from zero continuously close to a value of $2$. As a first indicator that the state goes from sub-Poissonian to a super-Poissonian distribution. Then, when the system goes through the crossover has an abrupt change from super-Poissonian to sub-Poissonian. Finally, for extremely larger density of excitations, the $g^2(0)$ goes toward $1$ from below. 

\begin{figure}[ht]
\centering
\includegraphics[width=0.49\textwidth]{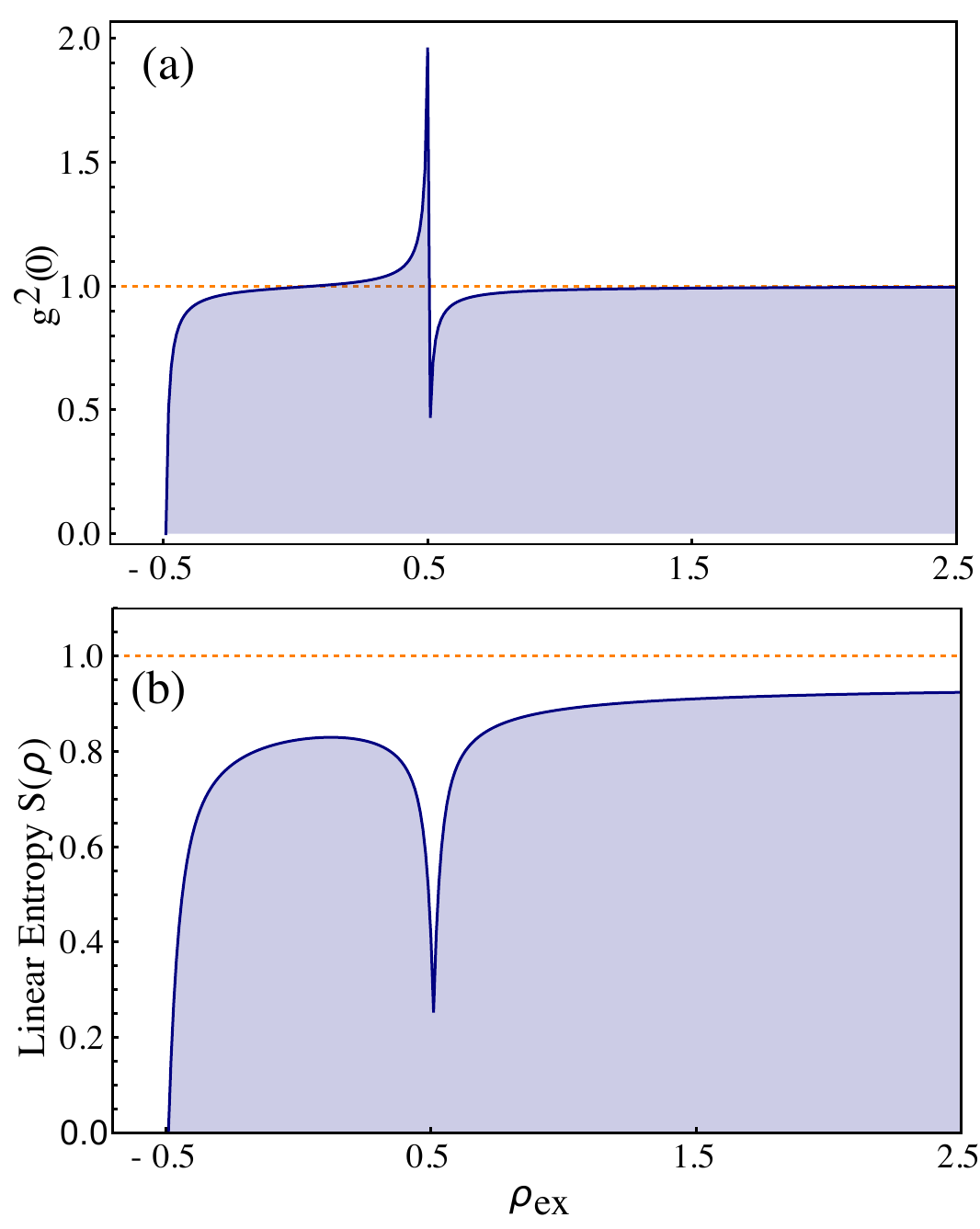}
\caption{Quantum correlations: (a) Second order correlation fucntion and  (b) Entanglement Entropy.}
\label{corr}
\end{figure}

In the following, we will concentrate our efforts examining entanglement properties in the different regimes. To do that, we calculate the linear entropy over the reduced density matrix, taking a partial trace,  either over matter or light. As it is shown in figure~\ref{corr}(b) the linear entropy increases from zero for $\rho_{\mathrm{ex}}=-0.5$ to a maximum value around $0.8$ at $\rho_{\mathrm{ex}}=0$ where it starts to decrease as the light loses its quantum behaviour. At $\rho_{\mathrm{ex}}=0.5$, just in the crossover, the system is close to be separable with almost a thermal state of light and a Fock state of matter ---a state with definite excitation--- this corresponds to a saturation of all the emitters at their maximum excitation. From this point, the entanglement increases gradually to its maximum value of $1$ as the density goes up.

Besides, we may get a more in-depth insight into the correlation structure of the physical system by analysing the first statistical moments, e.g., mean, variance, skewness, and kurtosis.  The definitions of these quantities were introduced in section~\ref{sec:formalism}. 

\begin{figure}[ht]
\centering
\includegraphics[width=0.50\textwidth]{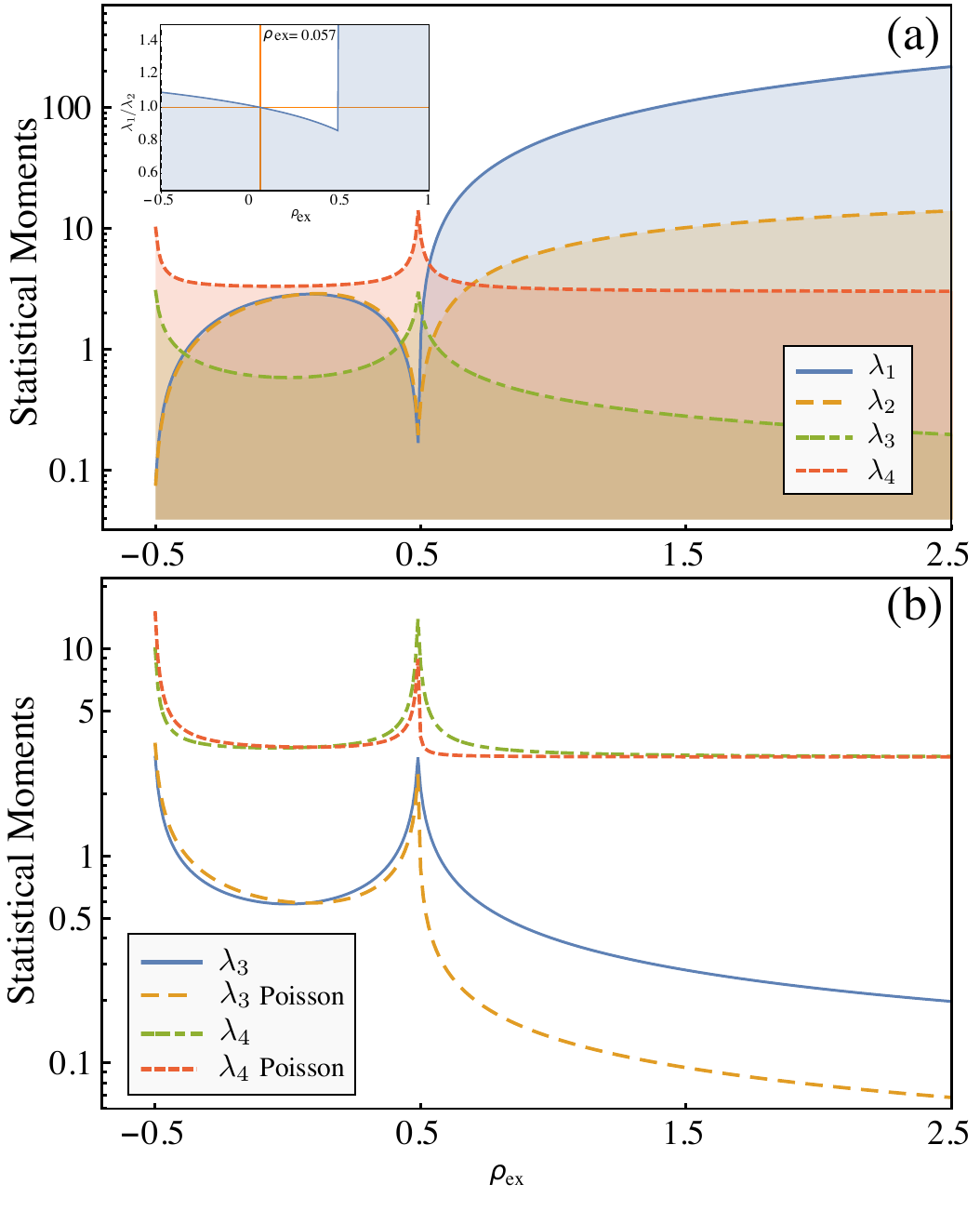}
\caption{Statistical moments: (a) first four statistical moments for the light field, i.e., $\lambda_1$ mean number of photons,  $\lambda_2$ variance, $\lambda_3$ skewness, and $\lambda_4$ kurtosis. In panel (b) we show the comparison with the statisticaa moments $\lambda_3=1/\sqrt{\lambda_1}$ and $\lambda_4=3+1/\lambda_1$ of a Poisson distribution.}
\label{fig:momentsl}
\end{figure}

\begin{figure}[ht]
\centering
\includegraphics[width=0.50\textwidth]{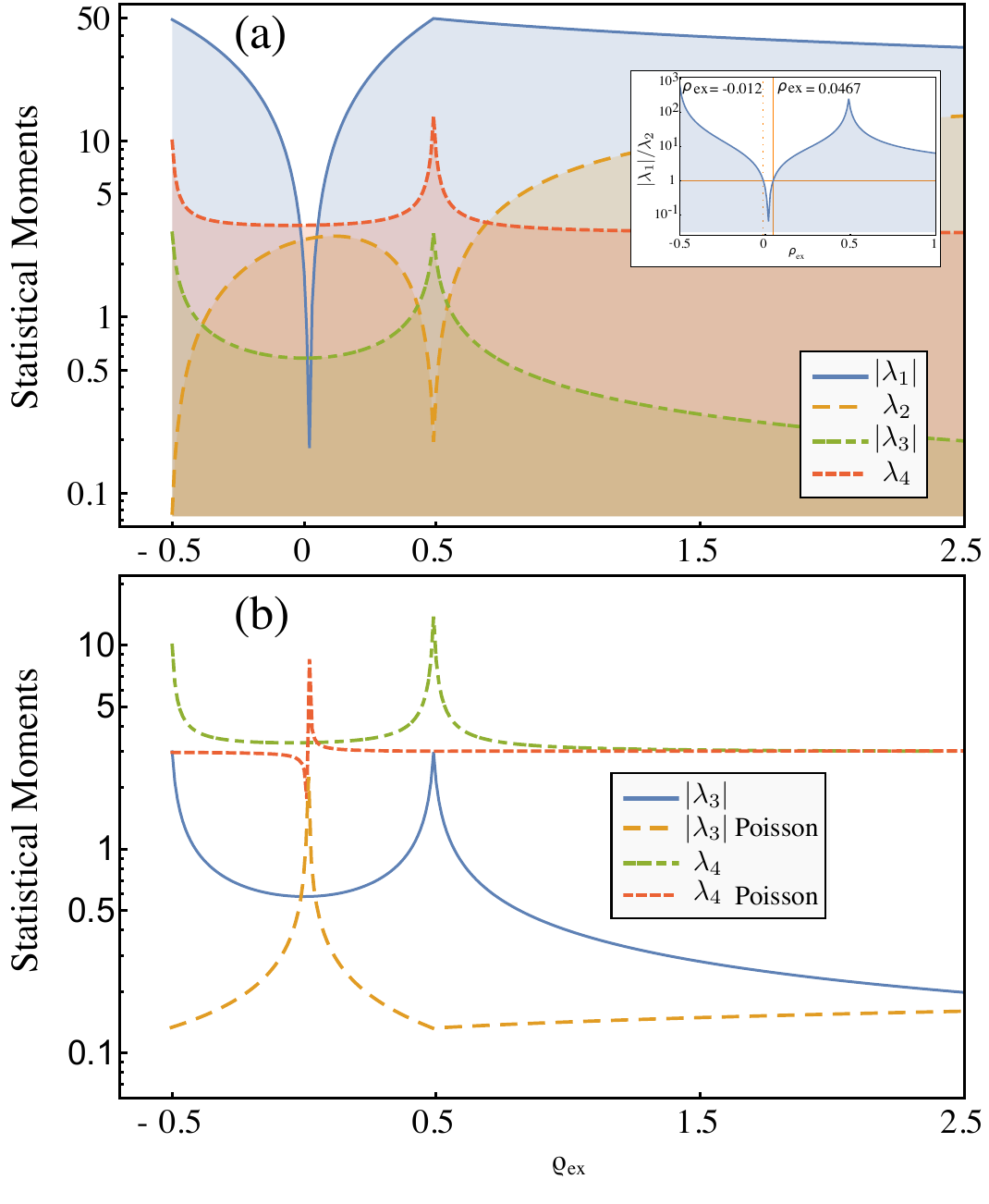}
\caption{Statistical moments: (a) first four statistical moments for the light field, i.e., $\lambda_1$ mean number of photons,  $\lambda_2$ variance, $\lambda_3$ skewness, and $\lambda_4$ kurtosis. In panel (b) we show the comparison with the statisticaa moments $\lambda_3=1/\sqrt{\lambda_1}$ and $\lambda_4=3+1/\lambda_1$ of a Poisson distribution.}
\label{fig:momentsm}
\end{figure}

We will centre around two aspects of the statistical moments: Their intrinsic relationships and a comparison with the actual statistical moments of a Poisson distribution.  Figure~\ref{fig:momentsl}(a) shows the first four statistical moments denoted as $\lambda_i$. As was intuited so far, for densities $\rho_{ex}$ between $-0.5$ and $0$, the variance stay under the mean of photons $\lambda_2<\lambda_1$. This fact confirms that in this region, the statistics is sub-Poissonian. This behaviour is smoothly interchange at  $\rho_{ex}=0.05703$ from where the statistics becomes super-Poissonian as is clear in the inset. This value does not coincide with the smooth interchange of statistic in the matter counterpart as we shall see later. At the crossover,  statistics undergoes an abrupt jump (see the inset of fig.~\ref{fig:momentsl}(a)]). It suddenly goes from super-Poissonian to sub-Poissonian. Contrary to the intuition, despite that the $g^2(0)$ goes to $1$ as the $\rho_{\mathrm{ex}}$ increases, the state of light is not a quantum coherent state. As it is clear from the figure, the statistics becomes strongly sub-Poissonian $\lambda_2\ll\lambda_1$. Figure~\ref{fig:momentsl}(2) reinforces this conclusion when we compare the actual moments with their counterpart of a Poisson distribution. Particularly, the skewness departs each other for large densities. On the other hand, the statistical moments for the matter subsystem is shown in figure~\ref{fig:momentsm}. In  panel (a) are shown the corresponding first four statistical moments. Here the mean excitation number can take negative values. Therefore, we depict its absolute value without normalising with the number of emitters $\mathrm{N}$. At first sight, the statistics of the assembly of emitters reminds sup-Poissonian for almost all densities; only in a very narrow region, it becomes super-Poissonian. As $\rho_{\mathrm{ex}}$ decreases from  $0.5$ the system reaches its smooth transition at $\rho_{\mathrm{ex}}=0.046892$, in other words, the coherence is reached by the light for a slightly high density; this is because the system is not in a resonance condition taking into account that the phase transition only appears for $\Delta>2$~\cite{eastham01a}. From figure~\ref{fig:momentsm}(b) is clear that the assembly's statistics does not compare with a Poisson distribution for any of the regions of the parameters. It remains sub-Poissonian, which is an expected result because of the sharp quantum character of the assembly.

\begin{figure}[ht]
\centering
\includegraphics[width=0.50\textwidth]{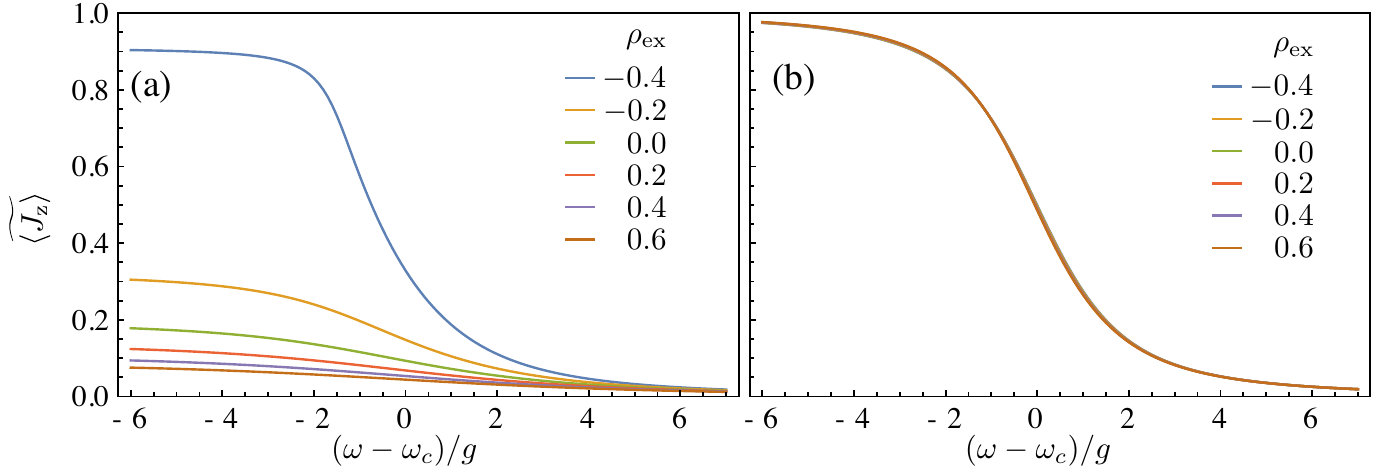}
\caption{Universal scaling law: (a) number of emitters $\mathrm{N}=10$, (b) number of emitters $\mathrm{N}=1000$. The population inversion scales universally with the density of excitations and following the factor  $1/(\rho_{\mathrm{ex}}+1/2)$.}
\label{fig:slaw}
\end{figure}

Even though the  population inversion $\braket{J_\mathrm{z}}$ of the assembly differs for the different excitation manifolds, it is possible to obtain a general criterion for establishing a universal scaling law in the system. We assess the occupation number as a function of the emitters energy $\omega=\omega_a$ for different values of excitation density $\rho_{\mathrm{ex}}=\{0.4$, $-0.2, 0.0, 0.2, 0.4, 0.6\}$ in both, the limit of few particles and the thermodynamic limit $\mathrm{N}\to\infty$. At few particles, see figure~\ref{fig:slaw}(a), we calculate the population inversion
$\widetilde{\braket{J_\mathrm{z}}}=(\braket{J_\mathrm{z}}+1/2)/(\rho_{\mathrm{ex}}+1/2)$. Only the excitation $\rho_{\mathrm{ex}}=-0.4$ goes to 1 for negative detuning; other excitations differ significantly from the rest of the excitations for all values of energy. On the other hand, as the number of emitters increases $\mathrm{N}=1000$ the value
of  $\widetilde{\braket{J_\mathrm{z}}}$ for all excitation densities collapse to the same curve, as it is shown in figure~\ref{fig:slaw}(b), because each one is divided with the universal scaling factor $1/(\rho_{\mathrm{ex}}+1/2)$ which defines a universal scaling law at thermodynamic limit.    

%%%%%%%
%%%%%%%

\section{Conclusions}
\label{conclusions}

Additionally, to the well-known behaviour of the BEC-QCP phase transition, we have complete the general picture by analysing the bipartite quantum correlations along with the structure of the statistics  for the photonic and matter subsystems. We may conclude that, before the crossover, the coherence is reached at two values of the density of excitations. The assembly of emitters reach the coherence at $\rho_{\mathrm{ex}}=0.04689$, just slightly before the photonic field  $\rho_{\mathrm{ex}}=0.05703$. Likewise, around the crossover, the system has two different behaviours. It undergoes an abrupt jump between them at  $\rho_{\mathrm{ex}}=0.5$ going from a super-Poissonian to sub-Poissonian structure of the statistical distributions of light. At this point, the entanglement goes from its minimum value. Contrary to the intuition, at very large densities, the statistics of light becomes strongly sub-Poissonian despite $g^2(0)\approx1$. Finally, for all the excitation manifolds, the systems follow and universal scaling law with the inverse of the excitation density $1/(\rho_{\mathrm{ex}}+1/2)$. This behaviour is clear from figure~\ref{fig:slaw} where the population inversion collapse to single behaviour with the former factor in the thermodynamic limit  $\mathrm{N}\to\infty$. As suggested above, the state of the whole system is an entangled state which is almost separable only just at the crossover when the assembly of emitters saturates.

\section*{Declaration of competing interest} 
The authors declare that they have no known competing financial interests or personal
relationships that could have appeared to influence the work reported in this paper.
\section*{Acknowledgments}
We gratefully acknowledge funding by COLCIENCIAS under the project ``Impact of phonon-assisted cavity feeding process on the effective light-matter coupling in quantum electrodynamics'', HERMES 47149. J.P.R.C. gratefully acknowledges support from the “Beca de Doctor- ados Nacionales de COLCIENCIAS 785”.

\bibliographystyle{elsarticle-num} 
\bibliography{Ref}

\end{document}